\documentclass[12pt]{article}
\usepackage[margin=2.5cm]{geometry}
\usepackage{longtable}
\usepackage{appendix}
\usepackage{subfigure}
\usepackage{amsmath}
\allowdisplaybreaks
\usepackage{color}
\usepackage{footnote}
\usepackage{algorithm}
\usepackage{algpseudocode}
\usepackage{algorithmicx}
\usepackage{multirow} 
\usepackage{booktabs}
\usepackage{graphicx}
\usepackage{amssymb}
\usepackage{amsbsy}
\usepackage{array}
\usepackage{longtable}
\usepackage{epstopdf}
\usepackage{pbox}
\usepackage{breqn}
\usepackage{mathrsfs}
\usepackage{multicol}
\usepackage{supertabular}
\usepackage{enumerate}
\usepackage[disable]{todonotes}
\usepackage{bbm}
\usepackage{bbold}
\usepackage{url}
\usepackage[justification=centering]{caption}

\usepackage{enumitem}
\setlist[enumerate]{noitemsep, topsep=0pt}
\usepackage[]{todonotes}

\newtheorem{thm}{Theorem}
\newtheorem{exm}{Example}
\newtheorem{prp}{Proposition}

\newcommand{\altk}{{h}}

\newcommand{\ntop}{\overline{n}} 
\newcommand{\nbot}{\underline{n}} 

\newcommand{\real}{\mathbb{R}} 

\newcommand{\I}{\mathcal{I}}

\newcommand{\Q}{\mathcal{Q}}

\newcommand{\N}{\mathcal{N}}

\newcommand{\Ell}{\mathcal{L}}

\newcommand{\bunderline}[1]{\underline{#1\mkern-2mu}\mkern4mu }

\begin{document}
\title{\bf Throughput-Improving Control of Highways\\Facing Stochastic Perturbations}
\author{Li Jin\thanks{Department of Civil and Urban Engineering, New York University, Brooklyn, NY, United States. Email: lijin@nyu.edu.},
  Alexander A. Kurzhanskiy\thanks{Institute of Transportation Studies, University of California, Berkeley, CA, United States.},
 and Saurabh Amin\thanks{Department of Civil and Environmental Engineering, Massachusetts Institute of Technology, Cambridge, MA, United States.}}
 \date{}
\maketitle


%
%
%

\begin{abstract}                          
In this paper, we study the problem of traffic management in highways facing stochastic perturbations. To model the macroscopic traffic flow under perturbations, we use cell-transmission model with Markovian capacities.  The decision variables are: (i) the admissible flow through each on-ramp, and (ii) whether individual on-ramps are metered to prioritize the mainline traffic or not. The objective is to maximize the throughput while ensuring that on-ramp queues remain bounded on average.
We develop a computational approach to solving this stability-constrained, throughput-maximization problem. 
We establish a mixed-integer linear program for throughput maximization and construct an algorithm that gives the optimal solution in particular settings.
We illustrate the performance benefits of the proposed approach through a computational study on a segment of Interstate 210 in California, USA. 
\end{abstract}


\section{Introduction}
We develop a control-theoretic approach to jointly determining, in a highway segment facing stochastic capacity perturbations, control inputs for managing traffic inflows through each on-ramps and the state of ramp meters, by considering the following problem:
\begin{align}
\max&\quad\mbox{throughput}\tag{P0}\label{eq_P0}\\
s.t.&\quad \mbox{every on-ramp queue is bounded on average}\nonumber&&
\end{align}
\todo{1.1, 1.13, 2.2}{\color{black}The decision variables in the above problem are (i) the closest traffic signals upstream to the highway on-ramps, which govern the admissible flow through each on-ramp, and (ii) the on/off state of the on-ramp controllers, which regulate the flows entering the mainline.}
\todo{1.4, 1.5}{\color{black}These decision variables determine the {\em controller configuration} of highway segment.}
\todo{1.7}{\color{black}The objective of \eqref{eq_P0} is to maximize the throughput, i.e. the total amount of inflow (demand) that is accepted and discharged by the highway under stochastic perturbations. This is roughly aligned with the objective of minimizing of travel time, since larger throughput typically implies shorter queues.}

Prior literature has considered related control
system design but mainly for deterministic-capacity settings \cite{como2016convexity,coogan16,papageorgiou2002freeway}, where traffic is modeled by deterministic macroscopic conservation laws (partial differential equations or their discretization) \cite{daganzo94,lighthill1955kinematic,richards1956shock}.
\todo{1.15}{\color{black}However, actual highway capacity is often subject to random fluctuation due to traffic incidents, breakdowns, or other latent features of highway infrastructure \cite{polus02,sumalee11}.}
In the presence of significant capacity perturbations, on-ramp queues that are nominally stable (in deterministic setting) can grow in an unbounded manner \cite{jin19tac}. 
This phenomenon highlights the relevance of developing a computational approach to solving \eqref{eq_P0}.
\todo{1.25}{\color{black}The following example illustrates why selecting the right controller configuration is essential for highways facing stochastic capacity perturbations.}

\begin{exm}
\footnotesize
\label{exm_1}
Consider the Interstate 210 (I210) section consisting of the mainline and an on-ramp (\todo{1.14, 1.16, 2.3}Fig.~\ref{fig_azusa}). We calibrated a cell transmission model with Markovian capacity using PeMS data\footnote{\scriptsize See \cite{varaiya09,jin2017calibration} for details on data source and calibration method.}.
\begin{figure}[hbt!]
\centering
\subfigure{
\centering
\includegraphics[trim=0 -3cm 0 0,clip, width=0.22\textwidth]{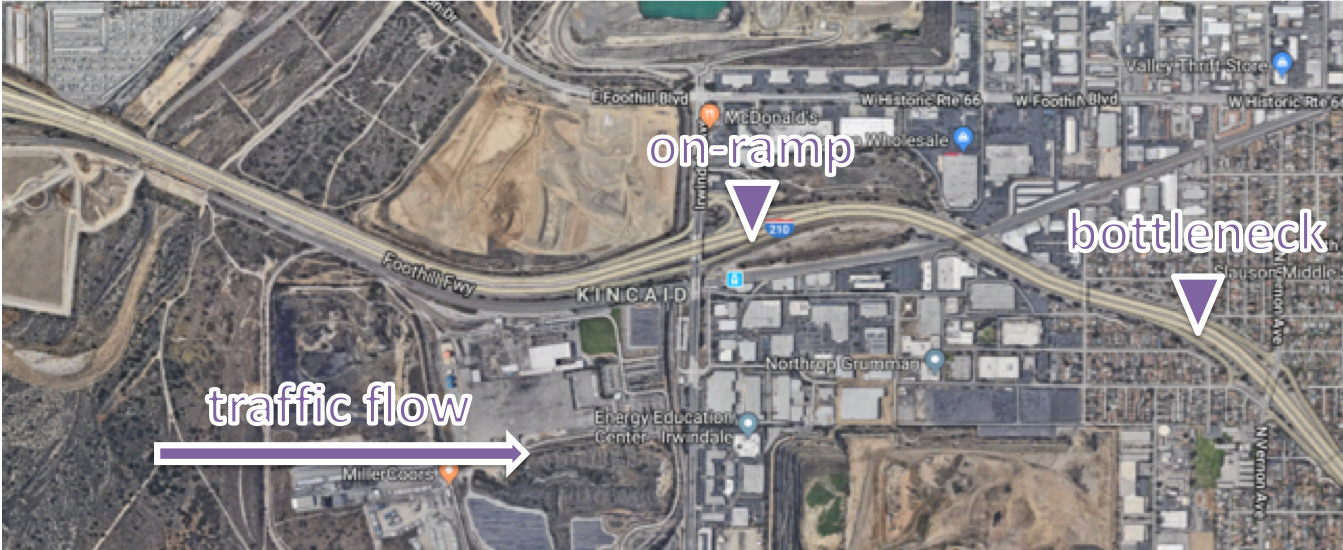}
}
\subfigure{
\centering
\includegraphics[width=0.22\textwidth]{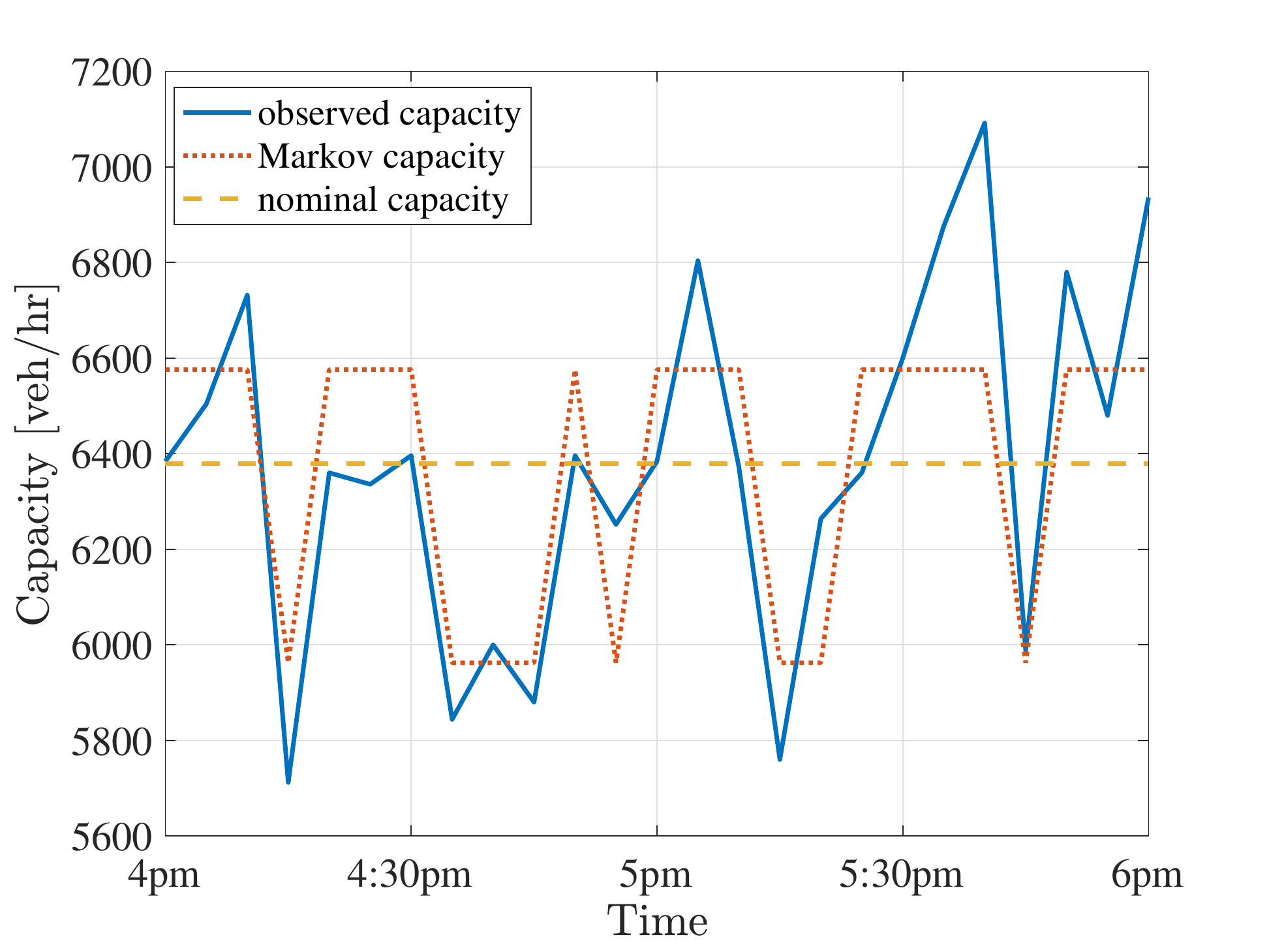}
}
\caption{A section of I210 (left) and {\color{black}the observed, modeled, nominal (fixed) capacities of the bottleneck (right).}}
\label{fig_azusa}
\end{figure}
Table~\ref{tab_time} shows the simulated travel times under three demand patterns and two control configurations, viz. metering off and metering on\footnote{\scriptsize In this example, a classical controller ALINEA \cite{papageorgiou2002freeway} is used for the ``metering on'' state.}. 
\todo{1.3, 2.3, 3.5}{\color{black}
A nominal scenario where capacity is constant (``nominal capacity'' in Fig.~\ref{fig_azusa}) is also considered. In this scenario, no congestion happens for demand patterns 1 and 2.}
{
\begin{table}[hbt]
\centering
\scriptsize
\caption{Simulated travel times various scenarios. ``Nominal'' means constant capacity without perturbations. ``Metering on/off'' is with perturbations.}
\label{tab_time}
\begin{tabular}{@{}cccc@{}}
\toprule
Demand pattern                       & 1            & 2              & 3            \\ \midrule
Mainline demand (veh/hr)                      & 5750 & 4700    & 5750       \\
On-ramp demand (veh/hr)               & 600  & 1650  & 1650       \\
Travel time (nominal, sec) & 83  & 75  & 758  \\
Travel time (metering off, sec) & 98  & 84  & 758  \\
Travel time (metering on, sec)           & 95       & 112         & 854         \\
\bottomrule
\end{tabular}
\end{table}
}
{\color{black}However, capacity perturbation has a significant impact on the travel time, and assuming a nominal capacity does not capture the impact.
In addition, although applying ramp metering slightly reduces travel time for relatively high mainline demand (pattern 1), it increases delay for relatively high on-ramp demand (pattern 2).}
Hence, the metering state needs to be carefully selected.
Furthermore, for the excessive demand (pattern 3), and the travel time is very large regardless of the metering state.
Thus, for certain high demand patterns, restricting the inflows admissible through the on-ramp may be preferable.
In conclusion, the selection of controller configuration in the presence of capacity perturbation depends on the demand pattern and the nature of capacity perturbation; (P0) captures this non-trivial decision problem.
\end{exm}

To formulate (P0) mathematically, in Section~\ref{sec_model}, we introduce a traffic flow model with Markovian capacities, based on the idea of the well-known cell transmission model (CTM) \cite{daganzo94}.
The dynamics is governed by (i) conservation law, (ii) flow-density relation (fundamental diagram \cite{daganzo94}), and (iii) stochastic capacity perturbtions.
\todo{1.12, 1.20}{\color{black}To some extent, our model is also relevant for capturing traffic flow dynamics with natural randomness \cite{sumalee11} and stochastic demand perturbations \cite{mehr2017stochastic}.}

In Section~\ref{sec_stability}, we establish an easy-to-check criterion for stability (boundedness) of the on-ramp queues under capacity perturbations.
These results allow us to formulate \eqref{eq_P0} as a mixed integer linear program (MILP).
Furthermore, for a typical class of perturbations (stationary incident hotspots), we develop an algorithm that constructs an optimal ramp metering configuration if the demand can be fully admitted.
We also present a computational case study using a calibrated simulation model to validate our approach.

The main contributions of this paper include (a) a stability criterion based on a set of linear inequalities, established by constructing an invariant set and verifying a Foster-Lyapunov stability condition, (b) a mixed-integer linear programming formulation for (P0) that can be solved using off-the-shelf solvers, (c) characterization of structure of optimal controller configuration for stationary incident hotspots, and (d) a simulation-based validation of the proposed approach.
\section{Highway Traffic under Stochastic Capacities}
\label{sec_model}

\begin{figure}[htb]
\centering
\includegraphics[width=0.35\textwidth]{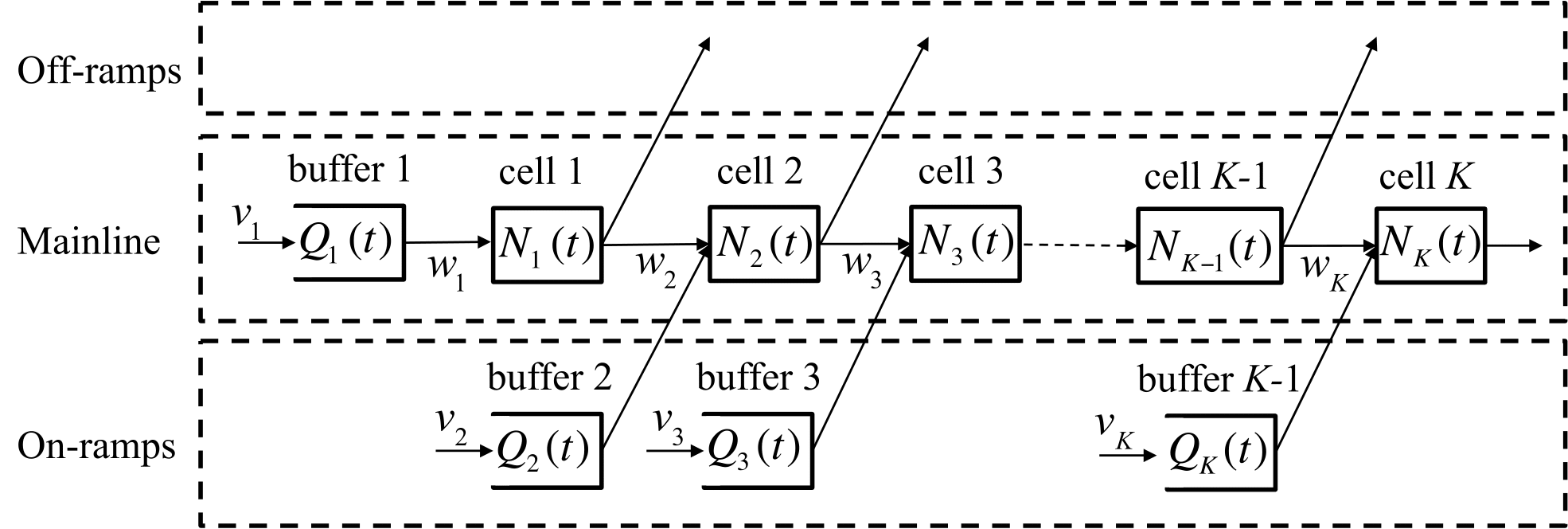}
\caption{A highway with $K$ cells and $K$ on-ramp buffers.}
\label{fig_ctm}
\end{figure}

We consider a highway with $K$ mainline \emph{cells}, $K$ on-ramp \emph{buffers}, and $K$ off-ramp exits, as shown in Fig.~\ref{fig_ctm}.
\todo{1.10}{\color{black}In practice, cell boundaries are determined according to locations of sensors, on-ramps, and off-ramps.}
For ease of presentation, we consider unit length for each cell; \todo{1.10}{\color{black}this assumption can be readily relaxed and does not affect the generality of our approach.}
Each cell $k$ has the following parameters: \emph{free-flow speed} $\alpha_k$ (km/hr), \emph{nominal capacity} ${\mathsf F_k}$ (veh/hr), \emph{congestion wave speed} $\beta_k$ (km/hr), and  \emph{jam (maximal) density} $n^\max_k$ (veh/km).
The $k$th buffer has a capacity ${\mathsf R_k}$ (veh/hr).

Let $x=(q,n)$ denote the \emph{continuous state} of the highway, where $q=[q_1\ q_2\ \cdots\ q_K]^T$ is the vector of lengths of the on-ramp queues and $n=[n_1\ n_2\ \cdots\ n_K]^T$ is the vector of traffic densities in mainline cells.
We assume that every buffer has an infinite size; this assumption enables us to focus on perturbation-induced congestion and excludes the impact of queue spillback to upstream roads. Thus, the set of permissible queue lengths is $\Q:=\real_{\ge0}^K$.
The set of traffic densities is $\N:=\prod_{k=1}^K[0,n^\max_k]$.

In our model, a controller \emph{configuration} is specified by $u=(v,w)$, where $v=[v_1\ v_2\ \cdots\ v_K]^T$ denotes the vector of \emph{inflows} (in veh/hr) admissible from upstream roads, and $w=[w_1\ w_2\ \cdots\ w_K]^T$ denotes the on-ramp controller state (metering ··on‘’ or metering ··off‘’). 
\todo{1.5}{\color{black}In many practical situations, transportation operators do not prefer to implement dynamic control configurations that vary frequently (e.g. every few minutes) \cite{kurzhanskiy2008aurora}. Hence, we only consider static controller configurations.}
The set of inflow vectors is $\mathcal V=\prod_{k=1}^K[0,d_k]$, where $d=[d_1\ d_2\ \cdots\ d_K]^T$ are the exogenously given (fixed) demand pattern.
\todo{1.2, 1.23}{\color{black}Indeed, actual demands are time-varying; however, we can reasonably use an average (constant) demand in designing a controller configuration that remains static for a period of several hours.}
Furthermore, we consider that any non-admitted demand is permanently rejected from the system and not redistributed to other locations.\footnote{\scriptsize Instead of $v_k\in[0,d_k]$ for each $k$, we can impose constraints such as $0\le\sum_{k=1}^Kv_k\le\sum_{k=1}^Kd_k$ to allow for redirection of demand. Such modification does not affect results in Section~\ref{sec_stability}.}
\todo{1.11}{\color{black}If a cell $k$ has no on-ramp, we impose $v_k=0$.}
The set of ramp controller states is $\mathcal W=\{0,1\}^K$; $w_k=1$ (resp. $w_k=0$) means that the $k$th on-ramp is in ``off'' (resp. ``on'') state.

\todo{1.25, 2.5}{\color{black}In our traffic flow model, the cell capacities stochastically vary over time according to a finite-state Markov process over a set of \emph{modes} denoted by $\mathcal I=\{0,1,2,\ldots,m\}$.
The inter-mode transition rates are $\{\nu_{ij};i,j\in\mathcal I\}$;}
\todo{1.17}{\color{black}
without loss of generality, we let $\nu_{ii}=0$ for all $i$.}
Every mode $i$ is associated with a vector of cell capacities $F(i)=[F_1(i)\ \cdots\ {F_K}(i)]^T$.
For ease of presentation, we assume that ${F_k}(i)\in\{{\mathsf F_k},{\mathsf F_k}-\Delta_k\}$ for each $k$ and for all $i$; that is, the capacity of the $k$th cell can only switch between two values ${\mathsf F_k}$ (nominal capacity) and ${\mathsf F_k}-\Delta_k$ (capacity reduced by $\Delta_k$).
We can still allow for deterministic capacity for any cell k by 
assuming that $\Delta_k=0$. Thus, $m\le 2^K$.
Let $I(t)$ be the mode at time $t$.
\todo{1.9}{\color{black}We assume that the Markov chain governing} \todo{2.12}{\color{black}the mode transition process $\{I(t);t\ge0\}$ is ergodic and admits with a unique (row) vector of steady-state probabilities $\mathsf p=[\mathsf p_0\ \mathsf p_1\ \cdots,\mathsf p_m]$;}
\todo{1.16}{\color{black}this essentially means that every capacity perturbation ends in finite time.}

We use $(i,x)=(i,q,n)$ to denote state variables and $(I(t),X(t))=(I(t),Q(t),N(t))$ to denote (hyrbid) stochastic process.
Following this convention, we let $Q_k(t)$ be the queue length in the $k$th buffer and ${N}_k(t)$ denote the {traffic density} in the $k$th cell at time $t$, and thus $Q(t)=[Q_1(t)\ \cdots\ Q_K(t)]^T\in\mathcal Q$ and $N(t)=[N_1(t)\ \cdots\ N_K(t)]^T\in\mathcal N$.

Next, we specify the traffic dynamics.
\todo{1.8}{\color{black}
For a fixed configuration $u=(v,w)\in[0,d]\times\{0,1\}^K$, the dynamics of  the on-ramp queues $Q(t)$ and traffic densities $N(t)$ can be written as follows:
\begin{subequations}
\begin{align}
&\dot Q(t)=G(I(t),X(t),u),\label{eq_qdot}\\
&\dot{N}(t)=H(I(t),X(t),u),
\label{eq_ndot}
\end{align}
\end{subequations}
where $G:\I\times(\Q\times\N)\times([0,d]\times\{0,1\}^K)\to\real^K$ and $H:\I\times(\Q\times\N)\times([0,d]\times\{0,1\}^K)\to\real^K$ are vector fields to be specified below.}

Following the classical CTM \cite{daganzo94}, for cell $k$, the {sending flow}~$S_k$ and the {receiving flow}~$T_k$ can be written as follows:
\begin{subequations}
\begin{align}
S_k(i,x)&:=\min\{\alpha_k n_k,{F_k}(i)\}\quad k=1,2,\ldots,K,\label{Eq_Sending}\\
T_k(x)&:=\beta_k(n^\max_k-{n}_k)\quad k=1,2,\ldots,K.\label{Eq_Receiving}
\end{align}
\label{Eq_SR}%
\end{subequations}
For $k=1,\ldots,K-1$, let $\rho_k\in(0,1]$ denote the fixed \emph{mainline ratio}, i.e. the fraction of traffic from cell $k$ entering cell $k+1$; the remaining traffic flow leaves the highway at the $k$th off-ramp.
\todo{1.18}{\color{black}Since the mainline ends at cell $K$, we set $\rho_K=0$; i.e. all outgoing flow from cell $K$ leaves the highway.}
In addition, we define
\begin{subequations}
\begin{align}
\rho_k^k&:=1\quad k=1,\ldots,K,\\
\rho_{k_1}^{k_2}&:=\prod_{\altk=k_1}^{k_2-1}\rho_\altk\quad 1\le k_1\le k_2-1,\; k_2=2,\ldots,K.
\end{align}
\label{Eq_betak1k2}%
\end{subequations}
The {sending flow} from the $k$th buffer is given by
\begin{align*}
D_k(x,u)=\left\{\begin{array}{ll}
\min\{v_k,{\mathsf R_k}\} & q_k=0,\\
{\mathsf R_k} & \mbox{o.w.}
\end{array}\right.
\quad k=1,\ldots,K.
\end{align*}
Thus, the \emph{flow} $r_k$ (resp. ${f_k}$) discharged by the $k$th buffer (resp. cell) is defined as follows:
\begin{subequations}
\begin{align}
&r_k(i,x,u):=\min\{D_k(x,u), (T_k(x)
-(1-w_{k})S_k(i,x))_+\}\nonumber\\
&\qquad k=1,\ldots,K,\label{eq_rk}\\
&{f_k}(i,x,u):=
\min\{S_k(i,x),(T_{k+1}(x)
-w_{k}D_{k+1}(x,u))_+\}\nonumber\\
&\qquad k=1,\ldots,K-1,\label{eq_fk-1}\\
&{f_K}(i,x,u):=S_K(i,x),\label{eq_fk}
\end{align}
\label{Eq_f(rho)}%
\end{subequations}
where $(\cdot)_+$ stands for the positive part.
One can see from \eqref{eq_rk}--\eqref{eq_fk} that if the $k$th on-ramp is metered (i.e. $w_k=0$), then mainline traffic is prioritized over the traffic on the $k$th on-ramp.



Then, the vector fields $G$ and $H$ in \eqref{eq_qdot}--\eqref{eq_ndot} follow from mass conservation:
\begin{subequations}
\begin{align}
&{G}_k(i,x,u):=v_k-r_k(i,x,u)\quad k=1,\ldots,K,\label{eq_G}\\
&{H}_1(i,x,u):=r_1(i,x,u)-f_1(i,x,u),\label{eq_H1}\\
&{H}_k(i,x,u):=\rho_{k-1}f_{k-1}(i,x,u)+r_k(i,x,u)-{f_k}(i,x,u)\nonumber\\
&\qquad k=2,\ldots,K.\label{eq_H}
\end{align}
\label{eq_GH}%
\end{subequations}
One can show that $G$ and $H$ are bounded and that $Q(t)$ and $N(t)$ are continuous in time $t$ \cite{jin19tac}. 

%

\section{Stability analysis}
\label{sec_stability}

In this section, we use the model defined above to concretize the boundedness constraint in (P0).
For a given configuration $u=(v,w)$, we say that the on-ramp queues are \emph{stable} if there exists $Z<\infty$ such that for each initial condition $(i,q,{n})\in \I\times \Q\times\N$,
\begin{align}
\limsup_{t\to\infty}\frac1t\int_{{s}=0}^t\mathrm E[|Q({s})|]d{s}\le Z.
\label{eq_stability}
\end{align}
\todo{1.21}{\color{black}Practically, bounded queues mean that one can expect a relatively small queue with high probability.}
We say that a demand vector $d$ is \emph{feasible} if there exists a stabilizing configuration $u=(v,w)$ such that $v=d$, and \emph{infeasible} otherwise.
\todo{1.23}{\color{black}Based on the above notion of stability, we define throughput $J$ as the inflow that is discharged through the highway:
$$
J(u):=\sum_{k=1}^Kv_k\quad\mbox{if \eqref{eq_stability} holds}.
$$
}%
Note that both $v$ and $w$ influence the stability condition \eqref{eq_stability}.
In the rest of this section, we establish a stability condition for the on-ramp queues.

For a given control input $u\in[0,d]\times\{0,1\}^K$, we construct an invariant set of the continuous state $x=(q,n)$ as follows:
$
{\mathcal M}(u):=
\bigcup_{{\theta}\in\prod_{k=1}^K[\bunderline q_k(u),\bar q_k(u)]}(\{\theta\}\times\prod_{k=1}^K[\nbot_k({\theta},u),\ntop_k(u)]),
$
where
\todo{2.12}
{\scriptsize
\begin{align*}
&\bunderline q_k(u):=\left\{\begin{array}{ll}
\infty & \mbox{if } v_k> {\mathsf R}_k\\
0 & \mbox{o.w.}
\end{array}\right.
\quad k=1,\ldots,K,\\
&\nbot_1({\theta},u):=\left\{\begin{array}{ll}
v_1/\alpha_1 & \mbox{if } {\theta}_1=0\\
\mathsf R_1/\alpha_1,& \mbox{o.w.}
\end{array}\right.\\
&\color{black}\nbot_k({\theta},u):=\left\{\begin{array}{ll}
\min\{\rho_{k-1}\nbot_{k-1}(q,u)+\frac{v_k}{\alpha_k},\frac{{\mathsf F}_k}{\alpha_k}\}
& \mbox{if } {\theta}_k=0\\
\min\{\rho_{k-1}\nbot_{k-1}(q,u)+\frac{{\mathsf R}_k}{\alpha_k},\frac{{\mathsf F}_k}{\alpha_k}\} &
\mbox{o.w.}
\end{array}\right.\\
&\color{black}\hspace{4.5cm} k=2,\ldots,K,\\
&\ntop_K(u):={n}^\max_K-\min_iF_K(i)/\beta_K,\\
&\ntop_k(u):=\min\Bigg\{\frac{{\mathsf F}_k}{\alpha_k},{n}^\max_k-\frac{\min_iF_k(i)}{\beta_k},{n}^\max_k\\
&\hspace{2.5cm}-\frac{{\beta_{k+1}(n^{\max}_{k+1}-\ntop_{k+1}(u))-v_{k+1}w_{k+1}}}{\rho_k\beta_k}\Bigg\}
\\
&\hspace{1in}
k=K-1,K-2,\ldots,1, \\
&\bar q_k(u):=\left\{\begin{array}{ll}
0 & \mbox{if } v_k\le{\mathsf R}_k\mbox{ and }v_k\le\beta_k(n^\max_k-\bar n_k(u))\\
&\;-(1-w_k)\rho_{k-1}\min\{\alpha_{k-1} \bar n_{k-1}(u),{\mathsf F}_k\}\\
\infty & \mbox{o.w.}
\end{array}\right.\\
&\hspace{5cm} k=1,\ldots,K.
\end{align*}}%
Intuitively, $\mathcal M(u)$ is a subset of the state space that the continuous state $Q(t)$ and $N(t)$ will eventually enter regardless of the initial condition; furthermore, once the state enters $\mathcal M(u)$, it will never leave.
\todo{2.7}{\color{black}The upper bounds $\bar n_k$ are computed assuming maximal downstream congestion, and the lower bounds $\underline n_k$ are computed assuming minimal upstream flow.}
\todo{2.13}{\color{black}For ease of presentation, we assume that $d$ is such that $\bar q_k=0$ if $w_k=1$; that is, if an on-ramp is not metered, then no queue will be formed there.
}
For the purpose of stability analysis, we partition ${\mathcal M}(u)$ into two subsets:
\begin{subequations}
\begin{align}
&{\mathcal M}_0(u):=\{(q,n)\in\mathcal M(u):q=0\},\label{eq_M0}\\
&{\mathcal M}_1(u):=\{(q,n)\in\mathcal M(u):q\neq0\},\label{eq_M1}
\end{align}
\end{subequations}
\todo{2.12}{\color{black}where $q=0$ means that $q$ is the vector of zeros.}%
We can also define the (finite) set of vertices of $\mathcal M_1(u)$ as follows:
$$
\mathcal V_1(u)=\cup_{\theta\in(\prod_{k=1}^K\{\underline q_k,\bar q_k\})\backslash\{0\}}
(\{\theta\}\times\{\underline n_k(\theta,u),\bar n(u)\}).
$$ 
In addition, let $e$ be a $K$-dimensional vector of 1's and
$$
C=\left[\begin{array}{ccccc}
\rho_1 & 1 & 0 & \cdots & 0\\
0 & 0 & 1 & \cdots & 0\\
\vdots & \vdots & \vdots & \ddots & \vdots\\
0 & 0 & 0 & \cdots & 1
\end{array}\right],
\quad
D=\left[\begin{array}{ccccc}
\rho_1 & 0 & \cdots &0 &  0\\
\vdots & \vdots & \ddots & \vdots & \vdots\\
0 & 0 & \cdots &\rho_{K-1} &  0\\
0 & 0 & 0 & \cdots & 1
\end{array}\right]
$$
Then, we can establish a sufficient condition as follows.

\begin{thm}
\label{thm_stable}
Consider a $K$-cell highway with a set of capacity perturbation modes $\mathcal I$ and with a demand vector $d\in\mathbb R_{\ge0}^K$. For a given control input $u=(v,w)\in[0,d]\times\{0,1\}^K$, the on-ramp queues are stable if there exist a symmetric matrix $A=(a_{kh})_{K\times K}$ satisfying
\begin{subequations}
\begin{align}
&a_{kh}\ge \rho_{k+1}a_{k+1,h}\quad k=1,\ldots,K-1,\ h=1,\ldots,K,\label{eq_akh}\\
&a_{K,K}>0\label{eq_aKK}
\end{align}
\end{subequations}
and a set of $K$-dimensional vectors $\{b^{(i)};i\in\mathcal I\}$, which jointly verify the following set of linear inequalities:
\begin{align}
&A\Big(CG(i,x,u)+DH(i,x,u)\Big)+\sum_{j\in\mathcal I}\nu_{ij}\left(b^{(j)}-b^{(i)}\right)\le-e\nonumber\\
&\qquad\forall(i,x)\in\mathcal I\times\mathcal{V}_1(u).\label{eq_stable}
\end{align}
\end{thm}

{\footnotesize \noindent{\bf Proof}
The proof relies on a Lyapunov approach; in particular, we establish stability by verifying the Foster-Lyapunov drift condition \cite{meyn93}:
\begin{align}
\Ell V\le-|q|+W
\quad\forall (i,q,n)\in\mathcal I\times\mathcal M(u)
\label{eq_drift}
\end{align}
where $\Ell$ is the infinitesimal generator (see \cite{meyn93} for definition) of the traffic flow dynamics and $W$ is some finite constant.
Consider the Lyapunov function $V:\mathcal I\times(\mathcal Q\times\mathcal N)\to\mathbb R_{\ge0}$:
\begin{align}
V(i,x):=\frac12(Cq+Dn)^TA(Cq+Dn)+(b^{(i)})^T(Cq+Dn).
\label{eq_V}
\end{align}
The main challenge for establishing \eqref{eq_drift} is that it must hold everywhere over $\mathcal M(u)$, which is a continuous set.
To address this, we show that it suffices to consider a finite set of states, which turns out to be $\mathcal V_1(u)$.
Suppose that there exist a matrix $A$ and vectors ${b}^{(i)}$ that satisfy the conditions in Theorem~\ref{thm_stable}.
For a given control input $u=(v,w)$, we have
\begin{align}
\Ell V=&
\Bigg((CG+DH)^T{A}+\sum_{i\in\mathcal I}\nu_{ij}\Big({b}^{(j)}-{b}^{(i)}\Big)^T\Bigg)
(Cq+Dn)\nonumber\\
&+(b^{(i)})^T
(CG+DH).
\label{eq_LViqn}
\end{align}
Since $G_k$ and $H_k$ are bounded; then we can define a constant
$
W:=\left(\max_{i,k}|b_k^{(i)}|\right)\sum_{k=1}^K\left(\mathsf R_k+\mathsf F_k\right)
$
and obtain that
\begin{align}
&(b^{(i)})^T(CG+DH)
{\le}
\left(\max_{i,x}|b^{(i)}_k|\right)\left(\max_{i,x}|CG+DH|\right)\nonumber\\
&{\le}
(\max_{i,x}|b^{(i)}_k|)
(\max_{i,x}(v_1-f_1+\sum_{k=2}^K(v_k-r_k
+f_{k-1}\nonumber\\
&+r_k-f_k)))\stackrel{\tiny\eqref{eq_rk}-\eqref{eq_fk}}{\le}
\left(\max_{i,k}|b_k^{(i)}|\right)\left(\sum_{k=1}^K\left(\mathsf R_k+\mathsf F_k\right)\right)\nonumber\\
&=W\quad
\forall (i,x)\in\mathcal I\times(\Q\times\N).
\label{eq_W}
\end{align}
The rest of this proof verifies the drift condition over ${\mathcal M}_0(u)$ and ${\mathcal M}_1(u)$ as defined in \eqref{eq_M0}--\eqref{eq_M1}:


For each $(q,n)\in{\mathcal M}_1(u)$ and for $h=1,\ldots,K$, note that
\begin{align*}
&(CG+DH)^T
\left[\begin{array}{ccc}{a}_{h,1} & \cdots &
{a}_{h,K}\end{array}\right]^T
=\rho_1{a}_{h,1}v_1+\sum_{k=2}^{K}{a}_{h,k-1}v_k\\
&-\sum_{k=1}^{K-1}({a}_{h,k}-\rho_{k+1}{a}_{h,k+1})(f_k+r_{k+1})-{a}_{h,K}f_K.
\end{align*}
\eqref{eq_akh} ensures that $a_{h,k}-\rho_{k+1}a_{h,k+1}\ge0$ for $k=1,\ldots,K-1$.
For each $q\neq 0$, $f_{k-1}+{r}_k$ is concave in $n$ over the box $\prod_{k=1}^K[\nbot_k(q),\ntop_k]$. Therefore, $f_{k-1}+{r}_k$ attains its minimum at one of the vertices of the box.
Then, we have
\begin{align*}
&{a}_{h,1}v_1+\sum_{k=2}^{K}{a}_{h,k-1}v_k
-\sum_{k=1}^{K-1}({a}_{h,k}-\rho_{k+1}{a}_{h,k+1})(f_k+r_{k+1})\\
&\qquad-{a}_{h,K}f_K
+\sum_{j\in\mathcal I}\nu_{ij}(b^{(j)}_k-b^{(i)}_k)\\
&\stackrel{\tiny\eqref{eq_stable}}\le {a}_{h,1}v_1+\sum_{k=2}^{K}{a}_{h,k-1}v_k-\min_{{(q,n)}\in{\mathcal V_1}}\Big(({a}_{h,k}-\rho_{k+1}{a}_{h,k+1})\\
&\quad\times(f_k+r_{k+1})+a_{h,K}f_k\Big)+\sum_{j\in\mathcal I}\nu_{ij}(b^{(j)}_k-b^{(i)}_k)\\
&
\le-1
\quad\forall(i,x)\in\mathcal I\times\mathcal M_1.
\end{align*}
Thus, from the above together with \eqref{eq_LViqn} and \eqref{eq_W}, we can obtain  that
$
\Ell V
\le-e^T(Cq+Dn)+W$
for all $(i,x)\in\mathcal I\times{\mathcal M}_1(u).
$
Recalling the definition of $C$, we have
$
\Ell V\le-|q|+W,
\quad\forall(i,x)\in\mathcal I\times{\mathcal M}_1(u).
$

Analogously, for each $(q,n)\in{\mathcal M}_0(u)$, it is not hard to show that
$\Ell V\le{Z_0}+W$.
Hence, letting $Z:=Z_0+W$, we can verify \eqref{eq_drift} and obtain the conclusion.
$\quad\blacksquare$}

When applying Theorem~\ref{thm_stable}, one can consider $A$ and $b^{(i)}$ as an unknown to be solved. 
Alternatively, by using insights about the traffic flow dynamics, we can use an explicit construction of $A$ which only leaves $b^{(i)}$ as unknowns. 
A simple construction is given by
$
a_{kh}=\gamma,
$
where $\gamma$ is a constant in $[1,\infty)$.
That is, the Lyapunov function $V$ as defined in \eqref{eq_V} penalizes the on-ramp queues and mainline traffic densities equally. 
An alternative construction is given by
$
a_{kh}=\gamma (K-k+1)(K-h+1),
$
where $\gamma$ is any constant in $[1,\infty)$.
This construction accounts for the spatial distribution of traffic: upstream congestion is penalized more than downstream congestion, and the Lyapunov function will decrease as traffic is discharged downstream. 
\section{Max-throughput problem}
\label{sec_formulation}

By Theorem~\ref{thm_stable}, we can concretely write (P0) as follows:
\begin{subequations}
\begin{align}
\max&\quad J(u)=\sum_{k=1}^Kv_k\label{eq_P1}\tag{P1}\\
s.t.&\quad A\Big(CG(i,x,u)+DH(i,x,u)\Big)+\sum_{j\in\mathcal I}\nu_{ij}\left(b^{(j)}-b^{(i)}\right)\nonumber\\
&\quad\le-e
\quad\forall(i,x)\in\mathcal I\times\mathcal V_1(u),\label{eq_GDH}\\
&\quad u\in[0,d]\times\{0,1\}^K.
\end{align}
\end{subequations}
The decision variables of \eqref{eq_P1} include the control input $u=(v,w)$, and Lyapunov function parameters $A$ and $b^{(i)}$. 
Since $G$ and $H$ as well as the states in $\mathcal V(u)$ are non-linear in $u$, even with a given matrix $A$, \eqref{eq_GDH} is non-linear in the $u$. 
Below is a reformulation (P1') of \eqref{eq_P1} such that
(i) when $A$ is considered as a variable, (P1') is a mixed integer bilinear program (MIBLP) with a linear objective function and bilinear constraints;
(ii) when $A$ is chosen to be fixed, (P1') is a mixed integer linear program (MILP).
{\scriptsize
\begin{subequations}
\begin{align}
\max&\ J(u)=\sum_{k=1}^K{v}_k\tag{P1'}\\
s.t.&\ \forall i\in\mathcal I,\ \forall{y}\in\{0,1\}^K\backslash\{0\}^K,\ \forall{z}\in\{0,1\}^K,\nonumber\\
&\;{a}_{h,1}{v}_1+\sum_{k=2}^{K}{a}_{h,k-1}{v}_k-\sum_{k=1}^{K-1}({a}_{h,k}-{a}_{h,k+1})
(\rho_k\tilde f_{k,y,z}^{(i)}\nonumber\\
&\qquad+\tilde r_{k+1,y,z}^{(i)})-{a}_{h,K}\tilde f_{k,y,z}^{(i)}
+\sum_{j\in\mathcal I}\nu_{ij}(b^{(j)}_h\nonumber\\
&\qquad-b^{(i)}_h)
\le-1+\sum_{k=1}^KM_1{y}_k{w}_k
\quad h=1,\ldots,K,\label{eq_drift2}
\\
&\; \tilde r_{k,y,z}^{(i)}\le {v}_k\quad\mbox{if }y_k=0,\
k=1,\ldots,K,\\
&\;\tilde r_{k,y,z}^{(i)}\le {\mathsf R}_k\quad
k=1,\ldots,K,\\
&\;\tilde f_{k,y,z}^{(i)}\le
\rho_{h}^k\left(\min_iF_h(i)\right)+\sum_{\ell=h+1}^{k-1}\rho_{\ell}^k{{v}_\ell}+\tilde r_{k,y,z}^{(i)}\nonumber\\
&\qquad\mbox{if }z_k=0,\
h=1,\ldots, k-1,\
 k=1,\ldots,K,\\
&\;\tilde f_{k,y,z}^{(i)}\le
\sum_{\ell=1}^{k-1}\rho_{\ell}^k{{v}_\ell}+\tilde r_{k,y,z}^{(i)}\ 
z_k=0,\
 k=1,\ldots,K,\\
&\;\tilde f_{k,y,z}^{(i)}\le F_k(i) \quad
k=1,\ldots,K,\\
&\;\tilde f_{k,y,z}^{(i)}\le
\frac{\tilde f_{k+1,y,z}^{(i)}-\tilde r_{k+1,y,z}^{(i)}}{\rho_k}\quad
\mbox{if }z_{k+1}=1,\nonumber\\
&\qquad k=1,\ldots,K-1,\\
&\; {v}_k
\le \tilde f_{k,y,z}^{(i)}-\rho_{k-1}\tilde f_{k-1,y,z}^{(i)}+M_2w_k
+M_2(1-{\xi}_k)
\nonumber\\
&\qquad k=2,\ldots,K,
\\
&\; {v}_k
\le \tilde f_{k,y,z}^{(i)}+M_2(1-w_k)+M_2(1-{\xi}_k)\
 k=2,\ldots,K,
\\
&a_{kh}\ge\rho_{k+1}{a}_{k+1,h}\quad
k=1,\ldots,K-1,
{a}_{Kh}\ge1\nonumber\\
&\quad h=1,\ldots,K,\
 v\in[0,d],\;w\in\{0,1\}^K,\;{\xi}\in\{0,1\}^K.
 \label{eq_uvw}
\end{align}
\end{subequations}}%
In the above, the decision variables include the controller configurations $v,w$, the Lyapunov function parameters $b^{(i)}$, and a set of auxiliary variables for the reformulation $\xi,\tilde f^{(i)},\tilde r^{(i)}$.
$M_1$ and $M_2$ are auxiliary for the big-$M$ constraints.
One can show that (P1) and (P1') are equivalent; the proof is omitted due to space constraint.
Note that (P1) and (P1') provide a lower bound for the optimal value of (P0), for the boundedness constraint in (P0) is replaced by sufficient condition for stability in (P1).

\begin{exm}
\footnotesize
This example shows how (P1') can be used to solve a practically relevant problem.
Consider a section of I210 near the on-ramp connected to I605, which is modeled as a two-cell highway (Fig.~\ref{fig_two} left).
\begin{figure}[htb]
\centering
\includegraphics[width=0.35\textwidth]{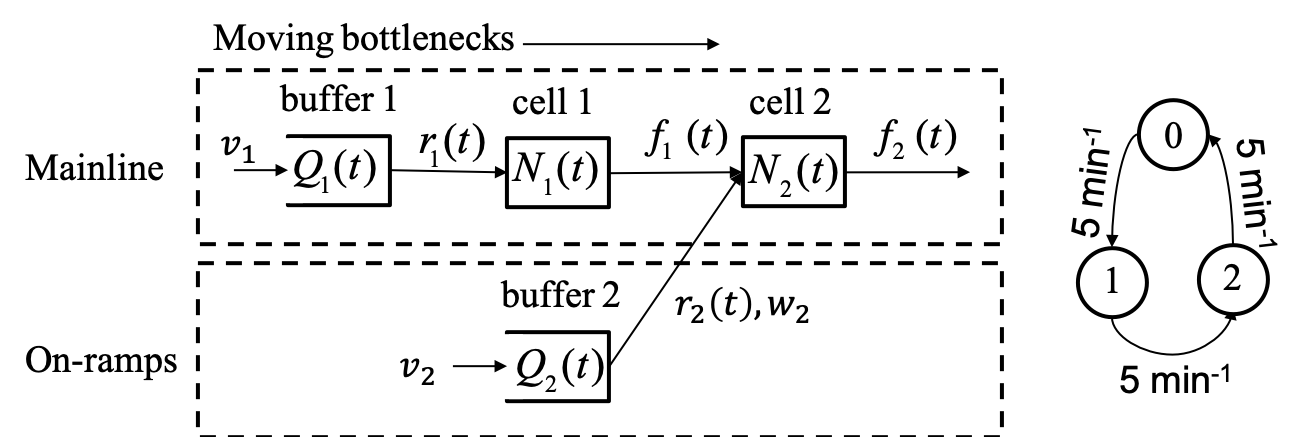}
\caption{A two-cell highway section (left) with HDV-induced capacity perturbations captured by a three-mode Markov chain (right).}
\label{fig_two}
\end{figure}
\todo{1.26}{\color{black}Consider the following practical scenario: heavy-duty vehicles (HDVs) randomly enter this section at a rate of per 5 min and spend on average 5 min in one cell. When an HDV is in a cell, the cell's capacity is reduced by 25\% (one lane). Such perturbations can be modeled as a three-mode Markov chain with $\mathcal I=\{0,1,2\}$ and $\nu_{01}=\nu_{12}=\nu_{20}=5$ per min (Fig.~\ref{fig_two} right).}
\todo{3.1}{\color{black}Fig.~\ref{fig_stable1} shows the feasible sets of the MIBLP and the MILP (with $a_{kh}=\gamma (K-k+1)(K-h+1)$), which are obtained by implementing Theorem 1 in YALMIP \cite{lofberg04}.
The meaning of each region is listed in Table~\ref{tab_stable1}.}
\begin{figure}[htb]
\centering
\includegraphics[width=0.3\textwidth]{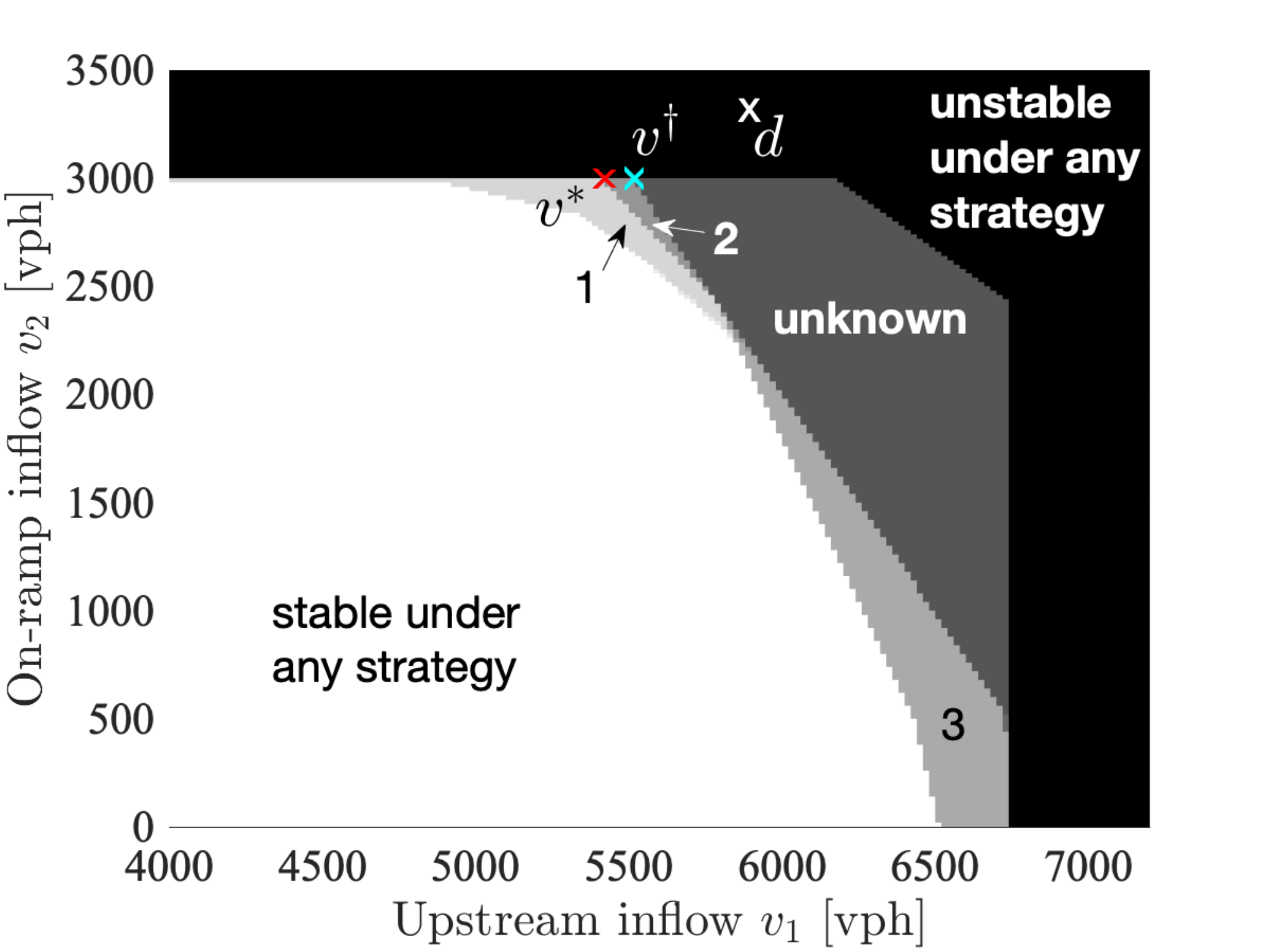}
\caption{Stability of various control configurations $(v,w)$.}
\label{fig_stable1}
\end{figure}
\begin{table}[hbt]
\scriptsize
\centering
\caption{Stability of various regions in Fig.~\ref{fig_stable1}.}
\label{tab_stable1}
\begin{tabular}{@{}ccccc@{}}
\toprule
Reg. & MIBLP & MILP & $w_2=0$ & $w_2=1$ \\ \midrule
1      & stable                     & stable                                  & unknown     &   stable         \\
2      & stable                     & unknown                                  & stable    &   unknown          \\
3      & stable                     & stable                                  & stable      &  unknown                             \\ \bottomrule
\end{tabular}
\end{table}
Actually Fig.~\ref{fig_stable1} can be used to graphically solve (P1') for this particular setting.
\todo{1.27}{\color{black}In particular, for a finite but infeasible demand $d$, $v^*$ (resp. $v^\dagger$) is an optimal solution to (P1') as an MILP (resp. MIBLP). Note that $v^\dagger$ is slightly higher than $v^*$. The optimal ramp state $w_2$ is 1, i.e. metering ``off''.}
\end{exm}

\subsection{Optimal solution for stationary incident hotspot}
\label{sub_structure}

\todo{1.26}{\color{black}
Suppose that for a $K$-cell highway, $\Delta_1=\Delta_2=\cdots\Delta_{K-1}=0$ and $\Delta_K>0$; cell $K$ is the only cell subject to capacity perturbations and called a stationary \emph{incident hotspot}.
}%
Analytical solution to (P1') for a general $K$-cell highway is not easy, even with a fixed $A$ matrix. 
However, we can analytically characterize the optimal solution for incident hotspots.

First, we start with the simple case where $K=2$.
For ease of presentation, we do not consider the impact of off-ramps and assume $\rho_1=1$.
Furthermore, we assume that ${\mathsf R}_1={\mathsf F}_1$.
\todo{1.3}{\color{black}
Then, the following result states that, for a feasible demand $d$, metering (resp. not metering) an on-ramp $k$ is optimal if $\mathsf R_2-d_2$, its capacity-to-demand margin, is greater (resp. smaller) than $\mathsf F_1-d_1$, the corresponding margin of the mainline}; an analogous result also holds for infeasible demands.
We call such structure of the optimal configurations the \emph{margin criterion}.

\begin{prp}
\label{prp_feasible1}
Consider a two-cell highway with a stationary incident hotspot. Suppose that $\rho_1=1$ and ${\mathsf F}_1={\mathsf R}_1$, and let $a_{kh}=\gamma>0$ for all $k,h$.
\begin{enumerate}
\item If the demand vector $d=[d_1\ d_2]^T$ is feasible, then
$([d_1\ d_2]^T,[1\ 1]^T)$ is an optimal solution to (P1) if
\begin{subequations}
\begin{align}
{\mathsf F}_1-d_1\ge {\mathsf R}_2-d_2,
\label{eq_F-d1>}
\end{align}
and $([d_1\ d_2]^T,[1\ 0]^T)$ is an optimal solution if
\begin{align}
{\mathsf F}_1-d_1\le {\mathsf R}_2-d_2.
\label{eq_F-d1<}
\end{align}
\end{subequations}

\item If $d_1=d_2=\infty$,
denote ${\bar F}_2=\mathsf p_0{\mathsf F}_2+\mathsf p_1({\mathsf F}_2-\Delta_2)$
and define
\begin{subequations}
\begin{align}
&\mathscr V_1^*:=\Big\{v\in[0,d]:
{v}_1+{v}_2={\bar F}_2,\
{\mathsf R}_2-v_2\ge {\mathsf F}_2\nonumber\\
&\qquad\qquad-{\bar F}_2,v_1< \min\{{\mathsf F}_1,{\mathsf F}_2-\Delta_2\}\Big\},\label{eq_U1}\\
&\mathscr V_2^*:=\Big\{v\in[0,d]:
{v}_1+{v}_2={\bar F}_2,\
{\mathsf F}_1-v_1\ge {\mathsf F}_2\nonumber\\
&\qquad\qquad-{\bar F}_2,v_2< \min\{{\mathsf R}_2,{\mathsf F}_2-\Delta_2\}\Big\}.
\end{align}
\end{subequations}
Then, every $(v,w)$ in the set $(\mathscr V_1^*\times\{[1\ 0]^T\})\cup(\mathscr V_2^*\times\{[1\ 1]^T\})$ is an optimal solution to (P1).

\end{enumerate}
\end{prp}

{\footnotesize
\noindent{\bf Proof}.
\emph{Part (1)}.
Consider a given $u=(v,w)\in[0,d]\times\{0,1\}^2$.
To obtain the conclusion, it suffices to show the following:
(i)  if $(v,w)=([d_1\ d_2]^T,[1\ 0]^T)$ satisfies \eqref{eq_drift2}--\eqref{eq_uvw} and if \eqref{eq_F-d1>} holds, then $(v,w')=([d_1\ d_2]^T,[1\ 1]^T)$ also satisfies \eqref{eq_drift2}--\eqref{eq_uvw};
(ii)  if $(v,w')=([d_1\ d_2]^T,[1\ 1]^T)$ satisfies \eqref{eq_drift2}--\eqref{eq_uvw} and if \eqref{eq_F-d1<} holds, then $(v,w)=([d_1\ d_2]^T,[1\ 0]^T)$ also satisfies \eqref{eq_drift2}--\eqref{eq_uvw}.
Here we only prove (i); (ii) can be analogously proved.

Suppose that $(v,w)=([d_1\ d_2]^T,[1\ 0]^T)$ satisfies \eqref{eq_drift2}--\eqref{eq_uvw} and that \eqref{eq_F-d1>} holds. 
If ${v}_1+{v}_2\le{\mathsf F}_2-\Delta_2$, then $\bar q_1=\bar q_2=0$ and $\mathcal M(u)$ is bounded regardless of $w_2$, and the proof is straightforward.
If ${v}_1+{v}_2> {\mathsf F}_2-\Delta_2$ and if ${v}_1\le{\mathsf F}_2-\Delta_2$, then, with $w_2=0$, \eqref{eq_drift2}--\eqref{eq_uvw} are equivalent to the following:
\begin{align}
&d_1+d_2-\mathsf p_0(d_1+{\mathsf R}_2)-\mathsf p_1\min\{d_1+{\mathsf R}_2, {\mathsf F}_2-\Delta_2)
\le-\delta,
\label{eq_d1+R2}
\end{align}
where $\mathsf p_0,\mathsf p_1$ are the steady-state probabilities of the Markov chain and $\delta$ is some positive number.
With $w_2=1$, \eqref{eq_drift2}--\eqref{eq_uvw} are implied by the following:
\begin{align}
&d_1+d_2-\mathsf p_0\min\{{\mathsf F}_1+d_2,d_1+{\mathsf R}_2\}-\mathsf p_1\min\{{\mathsf F}_1+d_2,
d_1\nonumber\\
&\qquad+{\mathsf R}_2, {\mathsf F}_2-\Delta_2) \le-\delta.
\label{eq_F1+d2}
\end{align}
If \eqref{eq_F-d1>} holds, then \eqref{eq_d1+R2} implies \eqref{eq_F1+d2}. Hence, $(v,w')=([d_1\ d_2]^T,[1\ 1]^T)$ also satisfies \eqref{eq_drift2}--\eqref{eq_uvw}.
If ${v}_1+{v}_2> {\mathsf F}_2-\Delta_2$ and if ${v}_1>{\mathsf F}_2-\Delta_2$, then $\bar q_1=\bar q_2=\infty$ with either $w_2=0$ or $w_2'=1$; thus, \eqref{eq_drift2}--\eqref{eq_uvw} are again equivalent to \eqref{eq_F1+d2}. Hence, $(v,w')=([d_1\ d_2]^T,[1\ 1]^T)$ also satisfies \eqref{eq_drift2}--\eqref{eq_uvw}.

\emph{Part (2).}
We only prove the optimality of solutions in $\mathscr V_1\times\{[1\ 0]^T\}$; $\mathscr V_2\times\{[1\ 1]^T\}$ can be analogously treated.
On the one hand, consider a $v^*\in\mathscr V_1^*$ and $w^*=[1\ 0]^T$. Then, we have
\begin{align}
&{v}_1^*+{v}_2^*-\mathsf p_0\min\{{v}_1^*+{\mathsf R}_2,{\mathsf F}_2\}-\mathsf p_1\{{v}_1^*+{\mathsf R}_2,{\mathsf F}_2-\Delta_2\}\nonumber\\
&{\le}{\bar F}_2-\mathsf p_0 {\mathsf F}_2-\mathsf p_1({\mathsf F}_2-\Delta_2)=0.
\label{eq_v1+v2}
\end{align}
One can then show that $(v,w)$ is a feasible solution to (P1).
On the other hand, ${v}_1^*+{v}_2^*\le{\bar F}_2$ is a necessary condition for stability, and ${\bar F}_2$ is the maximum throughput that can be achieved.
%
\hfill$\quad\blacksquare$}

\todo{1.25, 2.8}{\color{black}
In fact, we can extend the margin criterion to multi-cell highways with a stationary incident hotspot.
Consider a highway with $K$ cells and a feasible demand $d$ (so $v^*=d$); only cell $K$ is subject to capacity perturbations of magnitude $\Delta_K$. 
Suppose that $A$ in (P1) is given by $a_{kh}=\rho_{k+1}^K\rho_{h+1}^K$ for $k\le K-1,h\le K-1$, $a_{Kk}=a_{kK}=\rho_{k+1}$ for $k\le K-1$, and $a_{KK}=1$.
Then, the optimal priorities given by (P1') can be obtained using Algorithm 1.
\begin{algorithm}[htb!]
\scriptsize
\caption{Solution to (P2) for incident hotspot}\label{euclid}
\begin{algorithmic}[1]
\State $\underline T_K\gets\mathsf F_K-\Delta_K-d_K$\Comment{initialize}
\State{$\underline T_k\gets(\underline T_{k+1}-d_{k+1})_+,\ k=K-1,K-2,\ldots,1$}
\State{$\underline f_1\gets F_1$}
\For{$k=2:K-1$}\Comment{iterate all cells}
\State{$\underline f_k\gets\min\{\rho_{k-1}\underline f_{k-1}+d_k,\sum_{j=1}^{k-1}\rho_j^kd_j+R_k\}$}
\If {$\sum_{j=1}^{k+1}\rho_j^kd_j<\underline T_{k+1}/\rho_k$}
\State{$\underline f_k\gets\mathsf F_k,\ w_k\gets0$}
\ElsIf{$R_k-d_k> \rho_{k-1}(\mathsf F_{k-1}-\underline f_{k-1})$}
\State{$w_k\gets 0$}
\If{$\sum_{j=1}^{k}\rho_j^kd_j\le T_k$}
\State{$\underline f_k\gets \sum_{j=1}^{k-1}\rho_j^kd_j+R_k$}
\EndIf
\Else 
\State{$w_k\gets 1,\ \underline f_k\gets \underline f_{k-1}+d_k$}
\EndIf
\EndFor\label{euclidendwhile}
\State \textbf{return} $w$\Comment{return solution}
\end{algorithmic}
\end{algorithm}
Note that lines 8 and 13 are analogous to \eqref{eq_F-d1>}--\eqref{eq_F-d1<} and can be viewed as a generalization of the margin criterion. The proof of optimality of Algorithm 1 is somewhat involved but it is a mechanical extension of the proof of Proposition~\ref{prp_feasible1}.}
\todo{2.10}{\color{black}
Algorithm 1 significantly reduces the computation complexity compared to (P1').
We now present an example that utilizes this result as follows.}

\begin{exm}
\footnotesize
Consider a 7-km stretch of I-210 East model calibrated with real data.
%
\begin{figure}[htb]
\centering
\includegraphics[width=0.25\textwidth]{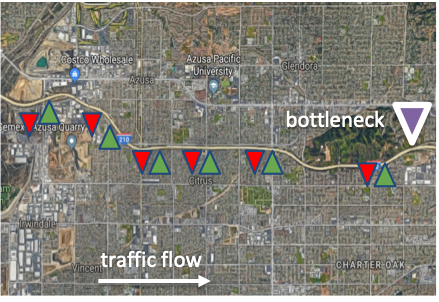}
\caption{\color{black}Configuration of the simulated highway section. Green (resp. red) triangles indicate on- (resp. off-)ramps.}
\label{fig_i210}
\end{figure}
%
\todo{3.3}{\color{black}
We consider a stationary incident hotspot near N. Azusa Ave. (``bottleneck'' in Fig.~\ref{fig_i210}).
The capacity of the hotspot switches between 6954 and 5968 veh/hr which, along with the transition rates, are obtained from PeMS \cite{varaiya09}.}
\todo{1.29, 3.4}{\color{black}
We ran three simulations in this setting, viz. (i) no on-ramp is metered, i.e. $w_k=1$ for all $k$, (ii) aggressive metering (every on-ramp is metered), i.e. $w_k=0$ for all $k$, and (iii) margin criterion, i.e. $w_k$ is determined by Algorithm 1.
}
\begin{figure}[hbt!]
\centering
\includegraphics[width=0.25\textwidth]{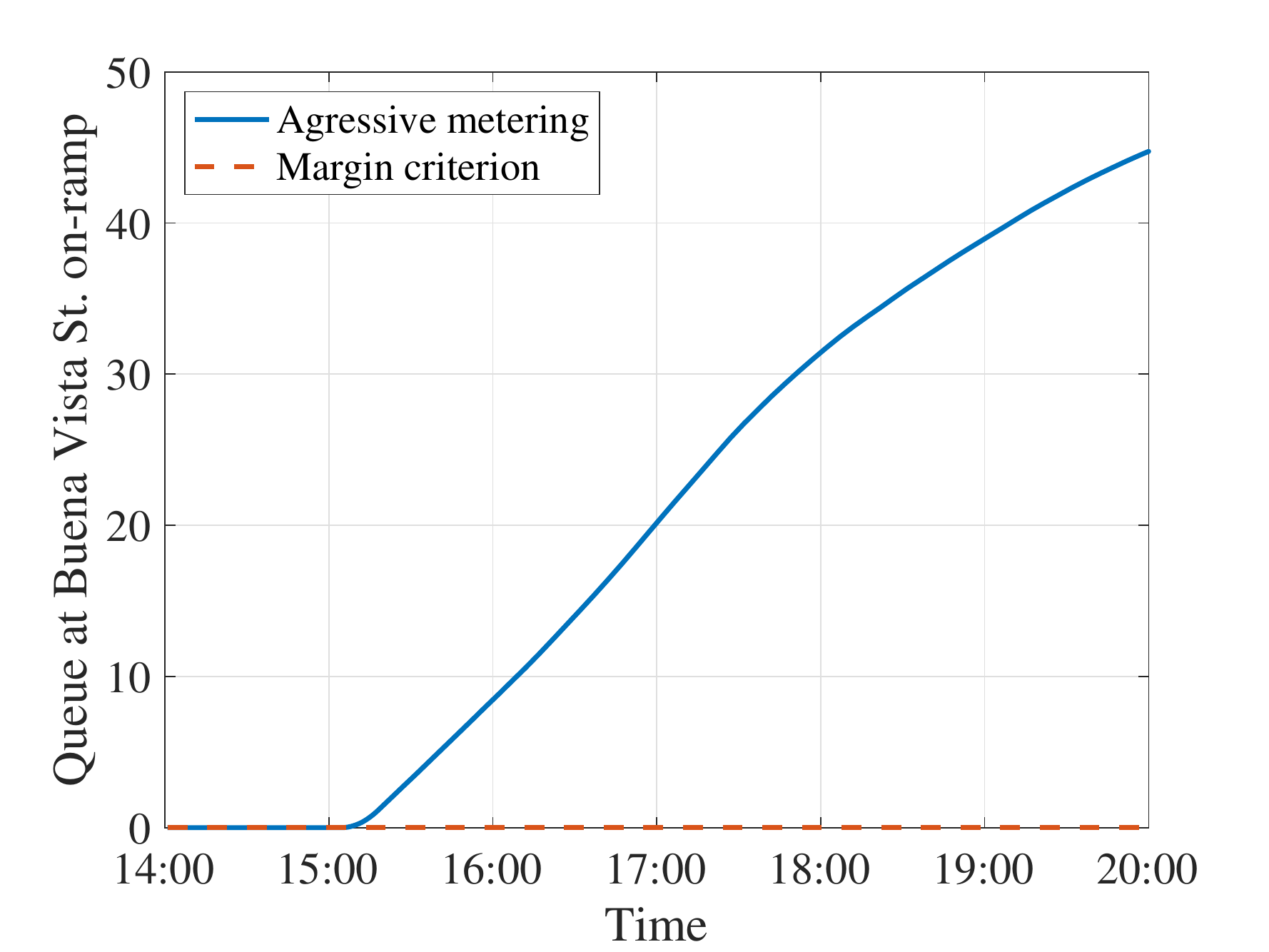}
\caption{Aggressive metering (i.e. metering every on-ramp) would lead to a long queue at an on-ramp (Buena Vista St.), while margin criterion does not meter this on-ramp and thus leads to zero queue (coinciding with horizontal axis).}
\label{Qk}
\end{figure}
The simulation results are as follows. First, as shown in Fig.~\ref{Qk}, aggressive metering leads to an on-ramp queue that grows very large; note that a queue of 50 vehicles spans approximately 0.5 km and would probably spill back to the upstream arterial road. Hence, aggressive metering is practically infeasible, and it is advantageous not to meter some on-ramps, e.g. the one indicated in Fig.~\ref{Qk}.
\todo{2.9}{\color{black}Second, the margin criterion prevents long on-ramp queues.
\begin{figure}[hbt!]
\centering
\subfigure{
\centering
\includegraphics[width=0.22\textwidth]{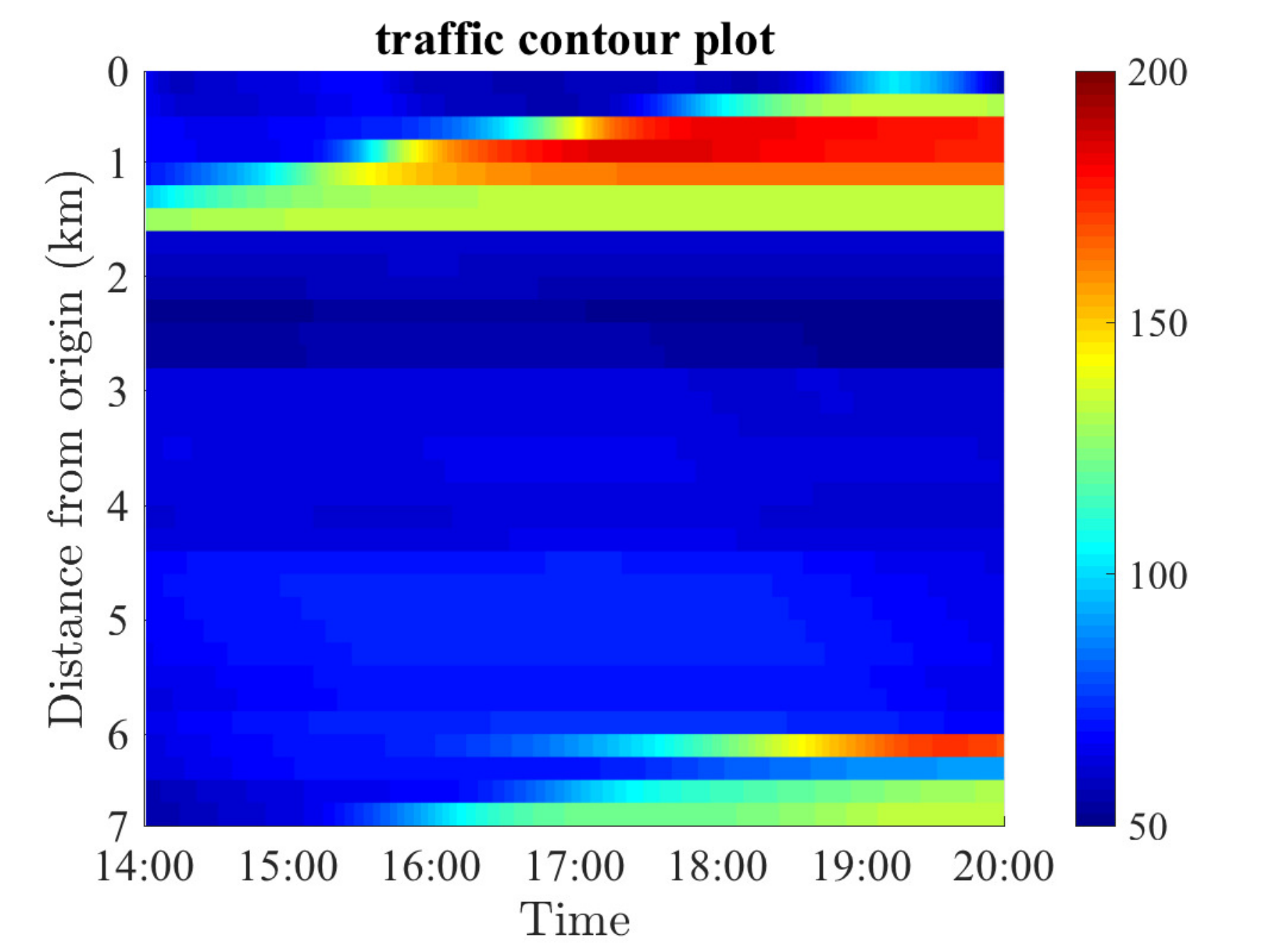}
\label{fig_i210_stat}
}
\subfigure{
\centering
\includegraphics[width=0.22\textwidth]{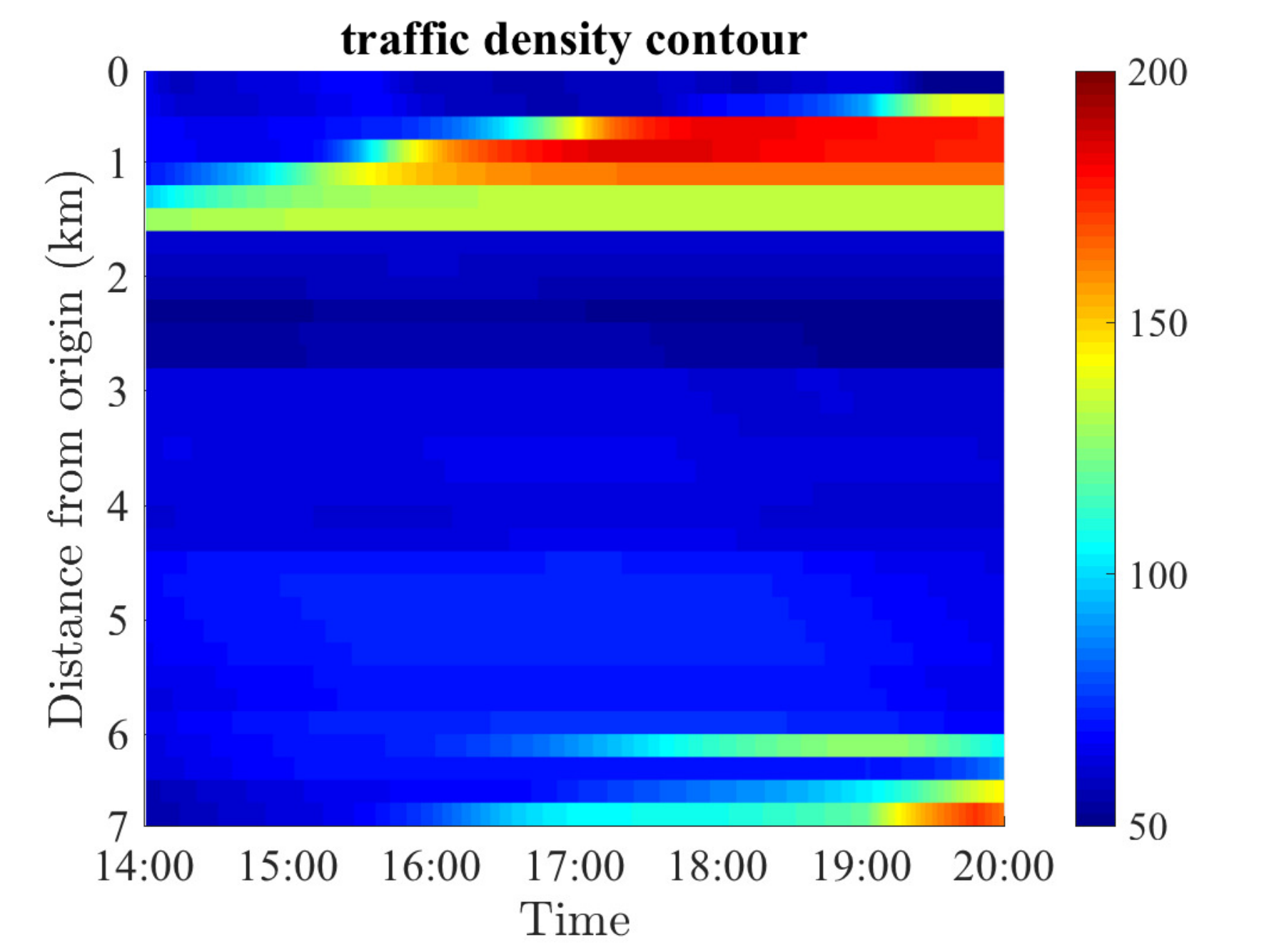}
\label{fig_i210_stat_mainline}
}
\caption{Traffic density (veh/km) contour under {\color{black}no metering (left)} and margin criterion (right).}
\label{fig_exm2}
\end{figure}
Importantly, margin criterion leads to 6\% less delay over one day and 1.2\% less throughput loss (i.e. higher throughput) during a peak hour (6-7pm) compared to no metering. (Here delay and throughput loss are computed against nominal free-flow conditions.)}
\end{exm}
\section{Concluding remarks}
\label{sec_conclude}

This paper develops a mixed integer linear formulation for determining the optimal controller configurations of highways facing stochastic capacity perturbations. For a particular class of perturbation (incident hotspots), the structure of optimal configurations is analytically characterized.
\todo{1.8, 2.2, 2.3, 3.2}{\color{black}
Potential future directions include using stability analysis results for solving to model-predictive control problems \cite{reilly2015adjoint,schmitt2018exact,yu2019traffic,zhang2017necessary} and
further statistical justification of the Markovian capacity model.
}

\bibliographystyle{plain}
\bibliography{Bibliography}
\end{document}